%% file: main.tex
\RequirePackage{amsmath}
\documentclass[runningheads]{llncs}
\usepackage[T1]{fontenc}
\DeclareFontShape{T1}{cmr}{m}{scit}{<->ssub*cmr/m/sc}{}
\usepackage{graphicx}
\usepackage{listings}
\usepackage{xcolor}
\usepackage{flushend} 
\usepackage{tikz}
\usetikzlibrary{shapes.geometric, arrows, positioning, matrix,arrows.meta,calc}

\definecolor{linkc}{rgb}{0.1,0.1,.8}
\definecolor{darkgreen}{rgb}{0,0.5,0}
\definecolor{midblue}{rgb}{0,0,0.7}
\definecolor{darkblue}{rgb}{0,0,0.5}
\definecolor{darkred}{rgb}{0.5,0,0}
\definecolor{titleblue}{rgb}{0.08,0,0.5}
\usepackage[bookmarks=false,colorlinks=true,citecolor=linkc,linkcolor=linkc,filecolor=linkc,urlcolor=linkc]{hyperref}
\usepackage{url}

\newtheorem{defn}{Definition}
\newtheorem{thm}{Theorem}
\newtheorem{examp}{Example}
\DeclareMathOperator{\shinv}{shinv}
\DeclareMathOperator{\shift}{shift}
\newcommand{\quo}{\mathbin{\mathrm{quo}}}
\newcommand{\rem}{\mathbin{\mathrm{rem}}}

\usepackage{cleveref}
\usepackage{pgfplots}
\usepackage{makecell}
\usepackage{multirow}
\usepackage{comment}

\usepackage{amssymb} 
\usepackage[ruled,vlined]{algorithm2e}
\SetKwInput{KwUse}{Uses}

\input{language}

\input{futhark}

\pgfplotsset{compat=1.18}

\newcommand{\mathword}[1]{\ensuremath{\text{\textit{#1}}}}

\newcommand{\spacetune}[1]{#1}
\newcommand{\stephenNote}[1]{}
\newcommand{\lstFigSize}{\scriptsize}
\newcommand{\lstDisplaySize}{\footnotesize}
\lstdefinestyle{lstFigStyle}{
   xleftmargin=2.5em,
   mathescape=true,
   escapechar=@,
   basicstyle=\lstFigSize\ttfamily,
   keywordstyle=\lstFigSize\color{midblue}\ttfamily\bfseries,
   commentstyle=\lstFigSize\slshape\color{darkgreen},
   numbers=left,
   numberstyle=\lstFigSize\rmfamily
}  
\lstdefinestyle{lstDisplayStyle}{
   xleftmargin=\parindent,
   mathescape=true,
   escapechar=@,
   basicstyle=\lstDisplaySize\ttfamily,
   keywordstyle=\lstDisplaySize\color{midblue}\ttfamily\bfseries,
   commentstyle=\lstDisplaySize\slshape\color{darkgreen},
   numbers=none
}  
\begin{document}
\title{On GPU Implementation for \mbox{Multi-Precision Integer Division} 
}
%
%

\author{Martin B. Marchioro\inst{1} \and Aske N. Raahauge\inst{1} \and Marc I. L\o{}venskjold\inst{1} \and \\ Cosmin E. Oancea\inst{1} \and Stephen M. Watt\inst{2}}
\authorrunning{Marchioro, Raahauge, L\o{}venskjold, Oancea, Watt}
%
%
\institute{DIKU, University of Copenhagen, Copenhagen 2100, Denmark
\email{martin.marchioro@gmail.com, aske.n.r@di.ku.dk,\\Marc.ivan95@gmail.com, cosmin.oancea@di.ku.dk} \orcidID{0000-0001-5421-6876}\\[3pt]
\and
Cheriton School of Computer Science, University of Waterloo, Canada\\
\email{smwatt@uwaterloo.ca} \orcidID{0000-0001-8303-4983}}
\maketitle
\begin{abstract}
This paper presents the issues arising in implementing a fast integer
division algorithm on general purpose GPUs. The algorithm uses a Newton
iteration based on the shifted inverse operation, keeping all arithmetic
in the integer domain and relying on data-parallel operators. The
principal contribution is an efficient GPU/\cuda{} implementation for
integer precisions from $2^{15}$ to $2^{18}$ -- sizes not supported by
\cgbn{} division. We propose algorithmic refinements, define a cost model
in terms of multiplications, build on prefix sums and previous work on
multi-precision multiplication, and present an evaluation showing
near-optimal performance relative to the model for the target precision.
\end{abstract}

\section{Introduction}

Multi-precision integer arithmetic is a basic component of computer
algebra, cryptography, exact scientific computation, and symbolic-numeric
software.  Its performance matters not only for isolated large
computations, but also for applications that require many independent
integer operations at the same precision.  This makes GPUs attractive:
they offer high arithmetic throughput and massive parallelism, provided
that the computation can be organized to keep data movement and
inter-thread communication under control.

Existing GPU support for multi-precision integers is strongest at
relatively small precisions, where one arithmetic instance can be mapped
to a small cooperative group of threads.  The \nvidia{} \cgbn{} library (Cooperative Groups Big Numbers) is an
important example of this approach~\cite{nvidialab:coopbignum}.  Such
libraries provide very high performance in their intended range, but they
do not cover all practically interesting sizes.  In particular, there is
a middle range of integer precisions---large enough that warp-level
methods become strained, but still small enough that a complete
arithmetic instance can fit within the fast memories of a single GPU
thread block.  This paper is concerned with this midsize regime.

In previous work we studied GPU implementations of multi-precision
addition and multiplication in this regime~\cite{midsize-add-mul}.  That
work showed that the classical algorithms, when scheduled carefully, can
be made effective on GPUs by assigning one multi-precision operation to a
\cuda{} block, keeping operands and intermediate values in registers or
shared memory, and minimizing global-memory traffic.  It also showed that
such algorithms can be expressed in a high-level data-parallel language
such as Futhark, although some low-level transformations needed for peak
performance remain beyond the current compiler.

Division is a more demanding operation.  It is not a simple local
operation on digits, and the usual high-performance approach is to reduce
division to multiplication by first computing an approximation to the
reciprocal of the divisor.  In conventional Newton iteration this
typically requires working in a domain where such reciprocals exist,
which can introduce multiple precision floating-point approximations and interactions between
different arithmetic domains.  For exact integer arithmetic this is
undesirable: the implementation must preserve exactness while still
exposing enough parallelism for the GPU.

The algorithm of Watt~\cite{watt2023} addresses this problem by replacing
the reciprocal with a whole shifted inverse.  Instead of computing
$1/v$, it computes an integer approximation to $B^h/v$, where $B$ is the
digit base and $h$ is an appropriate precision.  Newton iteration can
then be formulated using integer multiplication and shift operations.
This is attractive for GPUs because the main operations are
data-parallel: multi-precision multiplication, addition, subtraction,
comparison, and shifts.  Moreover, the algorithm is parameterized by the
multiplication method, so it can in principle benefit from either
classical multiplication or faster multiplication algorithms.

\enlargethispage{\baselineskip}

This paper investigates how this shifted-inverse division algorithm can be implemented efficiently on GPUs for midsize integers.  
Our present implementation is written directly in \cuda{} rather than generated from Futhark.  This is intentional: the goal is to identify the low-level scheduling, storage, and specialization issues that a high-level compiler would ultimately have to handle.  In particular, division stresses the compiler more than addition or multiplication because the Newton refinement uses operations whose effective precisions change during the iteration.
The main contributions of the paper are as follows:\vspace{-0.5ex}

\begin{itemize}
\item We present a \cuda{} implementation of multi-precision integer division based on the whole-shifted-inverse algorithm of Watt~\cite{watt2023}, targeting integer precisions from $2^{15}$ to $2^{18}$ bits.

\item We describe implementation refinements needed to make the algorithm robust in an unsigned-integer setting, including explicit sign handling in the close-product computation and quotient correction when the computed shifted inverse may overestimate by one.

\item We give a cost model for the implementation in terms of the number
of full multi-precision multiplications required.  For the classical
multiplication used here, the model predicts that the full division
operation should require at least five and at most seven full
multiplication costs.

\item We show how the supporting operations---shifts, comparisons, subtractions, close products, and variable-size multiplications---are mapped to \cuda{} blocks using registers and shared memory.

\item We evaluate the implementation on an \nvidia{} A100 GPU and compare it with \cgbn's corresponding operation.  The results show that, at the three of the larger precisions tested, the measured division time is between $6.77-6.94\times$ slower than one full multiplication, which is reasonably close to the lower bound of five multiplications predicted by the cost model, while also covering precision ranges not supported by \cgbn{} division in our experiments.
\end{itemize}

The rest of the paper is organized as follows.
Section~\ref{sec:division-alg} reviews the shifted-inverse division
algorithm, including the refinements needed for the implementation and
the resulting multiplication-based cost model.
Section~\ref{sec:gpu-implem} describes the \cuda{} implementation,
focusing on the use of registers, shared memory, scans, shifts,
subtraction, and variable-size multiplication.
Section~\ref{sec:eval} presents the experimental evaluation and compares
the implementation with \cgbn{} where the corresponding operations are
supported.
Section~\ref{sec:rel-work} discusses related work on GPU
multi-precision arithmetic and high-level data-parallel programming.
Section~\ref{sec:conclusion} summarizes the results and outlines future
directions, including clipped products and compiler support for this
class of exact arithmetic kernels.


\section{Division Algorithm}
\label{sec:division-alg}


\subsection{Intuition}

Parallel implementations for division of multi-precision integers typically rely on Newton's method to compute the reciprocal of the divisor, by applying it to function $f(x)=1/x-v=0$, where $v$ is the divisor and $x=\frac{1}{v}$ its reciprocal: 
%
\begin{equation}
x_{(i+1)}=x_{(i)} - \frac{f(x_{(i)})}{f'(x_{(i)})}=x_{(i)}+x_{(i)} \cdot \left(1-v\cdot x_{(i)}\right)
\label{eq:newton}
\end{equation}
However, Newton's method in its general form requires working in a related domain where the reciprocal exists. This can lead to a complex library structure in which the arithmetic domains are interdependent. As well, internal floating point representation can lead to potential loss of precision and overhead when converting between domains.
The algorithm proposed by Watt~\cite{watt2023} addresses these shortcomings and allows computation to be carried out in the integral domain, essentially by applying the Newton method to $f(x)=B^h/x-v=0$, where $B$ is the base of multi-precision integer $v$ and $h$ its precision, i.e., $B^h \geq v$. This computes the ``whole shifted inverse'' $x = \lfloor B^h/v \rfloor$ with the Newton iteration: 
%
\begin{equation}
w_{(i+1)} = w_{(i)}+\left\lfloor w_{(i)} \cdot (B^h-v \cdot w_{(i)}) \cdot B^{-h}) \right\rfloor, \quad w_{(i)} \in \mathbb{Z}
\label{eq:tmp}
\end{equation}
\Cref{eq:tmp} provides the intuition that the resulting Newton iteration can be written in terms of multiplications and efficient shift operations: 
\begin{equation} 
w_{(i+1)} = w_{(i)}+\shift_{-h}(\shift_h(w_{(i)})-v \cdot w_{(i)}^2), \quad w_{(i)} \in \mathbb{Z}
\label{eq:newton_shift}
\end{equation}
i.e., 
(1) shift operations are used to scale the input to a degree that captures all the necessary information from the fractional counterpart,
(2) then computations are performed at this higher scale, and
(3) a corresponding inverse shift is applied at the end to restore the 
original magnitude, avoiding potential loss of information. 
The shift and shift inverse operations are formally defined as:
%
\begin{defn}{(Whole shift and shifted inverse in $\mathbb{Z}$)} \\[0.5em]
Let $B > 1$ be an integer base. For integers $n$, $u$, and $v \neq 0$, with $n \geq 0$, the base-$B$ whole $n$-shift of $u$ and the base-$B$ $n$-shifted inverse of $v$ are defined as\vspace{-2ex}

\begin{equation*}
\shift_{n,B}(u) = \lfloor uB^n \rfloor
\qquad\qquad
\shinv_{n,B}(v) = \left\lfloor \frac{B^n}{v} \right\rfloor
\end{equation*}\vspace{-2ex}

When $B$ is clear from context, we write $\shift_n(u)$ and $\shinv_n(v)$.
\label{def:shiftshinv}
\end{defn}
When $n \geq 0$, $\shift_{n,B}(u)$ corresponds to integer multiplication, i.e., $u \cdot B^n$. When $n < 0$ it is instead a specialized quotient operation, with $u$ as dividend and $B^n$ as divisor. Using a multi-precision integer representation, this is equivalent to an arithmetic shift, e.g. $\shift_1([1,2,3])=[0,1,2]$ and $\shift_{-1}([1,2,3])=[2,3,0]$. 

In contrast, the whole shifted inverse $\shinv_n(v)$ corresponds to a specialized reciprocal, i.e., a reciprocal that has been shifted into our domain, e.g., $\shinv_3(8)=\shift_3(0.125)=125$.

\begin{thm}{(Quotient by whole shifted inverse in $\mathbb{Z}$)} \\[0.5em]
Given two positive integers $u$ and $v$, with $u \leq B^h$, we have:\vspace{-2ex}

\[
u \quo v \ \equiv \ \shift_{-h}~(~u \cdot \shinv_{h}(v)~) ~+~ \delta, \quad \delta \in \{0, 1\}.
\]
\label{thm:quo_by_shinv}
\end{thm}\vspace{-6ex}

Theorem~\ref{thm:quo_by_shinv} shows how these operations are used to derive the quotient:
applying a reverse shift on the result of multiplying $u$ with the whole shifted inverse of $v$ produces a result that is at most one unit away from the correct quotient.
%
%

\begin{examp}
\begin{rm}
Let $u = 314159265358979$, $v = 27183$ and find $q$ such that $u = q \times v + r$ for $0 \le r < v$. For $B=10$, $h = 15$ since $B^{14} < u < B^{15}$.
The iteration is given by equation~\eqref{eq:newton_shift}.
Start with initial guess $w_0 = 30000000000$. Then $ 30000000000 \rightarrow 35535300000  \rightarrow 36745061624 \rightarrow 36787648778 \rightarrow 36787698193 = \mathrm{shinv}_h(v)$ and $ q = \lfloor B^{-h}(u \times \mathrm{shinv}_h v \rfloor = 11557196238$.
\end{rm}
\end{examp}
\begin{examp}
\begin{rm}
Let $u = 726319138718412$, $v = 27183$.
Again, let $B = 10$, $h=15$ and initial guess
$w_0 = 30000000000$. Then
$30000000000 \rightarrow 35535300000 \rightarrow 36745061624
\rightarrow 36787648778 \rightarrow 36787698193
= \mathrm{shinv}_h(v)$ and
$q_0 = \lfloor B^{-h}(u \times \mathrm{shinv}_h v) \rfloor
= 26719609266$.  Since $u - q_0 v = 40734 \ge v$, the correction
$\delta = 1$ is needed, giving $q = q_0 + 1 = 26719609267$.
\end{rm}
\end{examp}




\subsection{Original Algorithm and New Revisions}
\label{subsec:revised-alg}

\input{incl/alg-shinv-revised}
\input{incl/alg-powdiff}

The pseudocode for computing the whole-shifted inverse is recounted in Algorithm~\ref{alg:shinv_revised}, which uses multi-precision addition,
subtraction and shift operations, and is defined in terms of a generic
multi-precision multiplication method, denoted by \textsc{mult}.
As well, it uses the \textsc{PowDiff} function, defined in Algorithm~\ref{alg-powdiff}.

\enlargethispage{\baselineskip}

The algorithm's correctness, fast convergence and work asymptotics are proven in~\cite{watt2023}. In short, the algorithm has asymptotic work equal to one multi-precision multiplication; this holds even when Strassen's $O(n\cdot \text{log}~n)$ algorithm~\cite{strassen1971schnelle} for multiplication is used. 
The algorithm consists of three main stages: special case handling, initial approximation and iterative refinement, which we summarize~below:\medskip

{\em Special Case Handling (lines \ref{g1-start}-\ref{g1-end})}  
ensures that the easy cases---corresponding to $B<v\leq B^h/2$---are handled, and that the base is sufficiently large ($B>=16$), such that the prerequisites for a good initial-value choice are met.
\medskip

{\em Initial Approximation (lines \ref{g2-start}-\ref{g2-end}).}
The original algorithm uses a three-digit approximation of the original $v$ value, namely
$V = \sum_{i=0}^{\ell} v_{k - \ell + i} \cdot B^i$, where $\ell = min(k,2)$. The shifted inverse is approximated to $w = \left(B^{2\ell} - V\right) \quo  V + 1$, which is more convenient and faster to compute than the equivalent $B^{2l} \quo  V$, albeit both take constant time. If sufficiently many digits are considered correct, i.e., $h - k \leq \ell$, then the original algorithm shifts $w$ to the appropriate magnitude and performs an early return, i.e., $\textsc{shift}_{h - k - \ell}(w)$, otherwise $w$ is refined.

We perform the revision shown in lines \ref{g2-start}-\ref{g2-end} of Algorithm~\ref{alg:shinv_revised} because there are rare corner cases in which the early return refers to an overestimated (unsafe) value (see ~\cite{msc-thesis} for an example). Instead, we always pass the initial approximation through \texttt{Refine}, which guarantees correctness. This also allows a less-precise initial approximation of $v$, namely $V = v_{k-1} + v_k\cdot B$ that uses only two (instead of three) digits, i.e., $\ell=2$, which is proven in~\cite{msc-thesis} to preserve fast convergence and is convenient since it promotes machine arithmetic.  
%
%

{\em Iterative Refinement} (functions \texttt{Refine}, \texttt{Step}, \texttt{PowDiff}).
The initial approximation is refined at least once with the fastest-convergence routine---named \texttt{Refine3} in~\cite{watt2023} and \texttt{Refine} here---that employs both shorter-iterates and divisor-prefixes techniques to achieve optimal work. 
%
%
%
\texttt{Refine} iteratively calls the \texttt{Step} function, which performs a single Newton iteration. \texttt{Step} invokes \texttt{PowDiff} (shown in Algorithm~\ref{alg-powdiff}), which computes $B^h - v \cdot w$ using the close-product strategy for improved efficiency. Since \textsc{PowDiff} can return negative integers and since our implementation assumes unsigned integers, we refine the original implementation to explicitly keep track of the integers' sign in \texttt{PowDiff} and \texttt{Step}. 


\input{incl/alg-div-new}

With divisor prefixes~\cite{watt2023}, overestimation can also arise
during refinement: when \texttt{Refine} uses only a prefix of $v$, rare
edge cases allow low digits to affect higher ones. It is
proven~\cite{msc-thesis} that Algorithm~\ref{alg:shinv_revised} can then
overestimate the shifted inverse by at most one, so
$\texttt{shift}_{-h}(u \cdot \texttt{shinv}_{h}(v))$ can be one below or
one above $u \quo v$. We therefore revise
Theorem~\ref{thm:quo_by_shinv} and amend Algorithm~\ref{alg:div_new}:


\begin{thm}\label{thm:quo_by_shinv_opt}{(Revised Quotient by shinv in $\mathbb{Z}$)} \\[0.5em]
Given two positive integers $u$ and $v$, with $u \leq B^h$, the following hold:\vspace{-1ex}
\begin{equation*}
\widehat\shinv_{h,B}(v) = \left\lfloor \frac{B^h}{v} \right\rfloor + \lambda, \quad \lambda \in \{0,1\}
\end{equation*}\vspace{-1ex}
\begin{equation*}
u \quo v = \shift_{-h}(u \cdot \widehat\shinv_h\, v) + \delta, \quad \delta \in \{-1, 0, 1\}
\end{equation*}
\end{thm}

\subsection{Algorithm Cost in Number of Full Multiplications}
\label{subsec:cost}

\enlargethispage{\baselineskip}

This section assumes multi-precision integers consisting of $M$ digits in base $B$ and approximates the cost of the division algorithm in terms of the minimal and maximal number of full multiplications that are performed with the classical/quadratic algorithm. We consider that a full multiplication is performed whenever the result requires to compute more than $M/2$ digits. In summary, the computation of the shifted inverse requires at least two and at most four full multiplications. Once the shifted inverse is known, the straightforward computation of the quotient and remainder shown in Algorithm~\ref{alg:div_new} requires:\vspace{-1ex}
\begin{itemize}
\item a full multiplication $v\cdot q$  at line~\ref{line:rem-init}, just before the quotient adjustment, and
\item a multiplication $u\cdot \shinv(v,h,B)$ at line~\ref{line:quot-init} in the computation that approximates the quotient. This multiplication has to be computed in double precision $2\cdot M$ because the result is shifted back by $h$ digits. Assuming  classical/quadratic multiplication, its cost is thus equal to {\em two} full multiplications.
\end{itemize}
It follows that the presented division algorithm requires {\em at least five and at most seven full multiplications}. The remainder of this section justifies the lower and upper bound of the cost of the whole shift inverse algorithm. Essentially, the loop inside the \texttt{Refine} function of Algorithm~\ref{alg:shinv_revised} exhibits at least one and at most two full multiplications inside \textsc{PowDiff} (called from \texttt{Step} and shown in Algorithm~\ref{alg-powdiff}) and similarly for the computation of \texttt{Step} excluding \textsc{PowDiff}.

A full multiplications inside \textsc{PowDiff} requires that $k > \frac{h}{2}$, where $k$ is the precision of (the original) $v$. The precision of the $v$ parameter of \textsc{PowDiff} is $\text{prec}_B(v) = \text{min}(2\cdot \ell_i, k)$, which is also a good approximation of $L$. It follows that a full multiplication is performed whenever $\lfloor \frac{h}{2} \rfloor < \text{prec}_{B}(v) \approx 2\cdot \ell_i$. However, we also know that the loop in \texttt{Refine} executes as long as $\ell_i < h - k$. Since $k > \frac{h}{2}$ it follows that $h-k \leq \lceil \frac{h}{2} \rceil$ and the loop terminates whenever $\ell_i$ reaches $\lceil \frac{h}{2} \rceil$.

The condition for performing a full multiplication was $\lfloor \frac{h}{2} \rfloor < 2\cdot \ell_i$, which is equivalent to $\lfloor \frac{h}{2} \rfloor - 1 < 2\cdot \ell_i - 1 \approx \ell_{i+1}$, since in most relevant cases the update formula for $\ell$ is $\ell_{i+1} = 2\cdot \ell_i - 1$. It follows that it is possible, albeit unlikely, to be in the case $\lfloor \frac{h}{2} \rfloor - 1 < \ell_{i+1} < \lceil \frac{h}{2} \rceil$ that requires the loop in \texttt{Refine} to execute another iteration performing a full multiplication.

Overall, the focus of an efficient implementation of this algorithm is to achieve a runtime close to that of five full multiplications, which critically requires that the computation outside said multiplication does not introduce~bottlenecks. 


\section{GPU Considerations}
\label{sec:gpu-implem}

\subsection{High-Level Rationale of the Implementation}

\enlargethispage{\baselineskip}
\enlargethispage{\baselineskip}

We report a \cuda{} implementation of the division algorithm of
\cref{sec:division-alg} for multi-precision integers whose computation
fits inside one \cuda{} block. Following prior work on addition and
multiplication~\cite{midsize-add-mul}, operands are copied once from
global memory to registers, results are copied back, and the remaining
computation uses fast memory. Intermediate arrays stay in registers when
possible and are materialized transiently in shared memory only when
communication or performance requires it, for example for overlapping
access in classical multiplication or as a staging buffer for coalesced
global-memory transfers.
%
In addition we apply classical techniques such as efficient sequentialization of excess parallelism to minimize inter-thread communication and thus maximize throughput. 

More detailed many-core models can account explicitly for memory
transactions, synchronization, occupancy, and parallelism
overheads~\cite{DBLP:conf/parco/HaqueMX15,MA2014202}.  Our cost model is
deliberately coarser: it uses the measured cost of our multi-precision
multiplication kernel as the architecture-aware unit of cost, since that
kernel uses the same representation, memory hierarchy, and block-level
execution strategy as the division kernel.
%
%
This section uses the following notation:\vspace{-1ex}
\begin{description}
    
    \item[\emph{uint:}] the word size representing a digit of the multi-precision integer; we use word sizes of $16$, $32$ or $64$ bits, since these are hardware supported. \smallskip

    \item[\emph{M:}] The total number of digits in the big integer. For example, an integer with $2^{17}$ bits could be represented using $M = 2048$ and a 64-bit word size. \smallskip
    
    \item[\emph{Sequentialization factor ($Q$):}] The amount of sequential work each thread performs. For simplicity, we assume that $Q$ evenly divides $M$.
\end{description}\smallskip

{\em Memory Limitations.}
Currently, our \textbf{\textsc{cuda}} implementation supports integer division on integers as large as $2^{18}$ bits. These sizes are limited by the maximum amount of shared memory available per \cuda{} block. For example, the \cuda{} implementation of classical multiplication~\cite{midsize-add-mul} requires manifesting both input arrays in shared memory, which sums up to $64$KB. Since the current practical maximal amount of shared memory per-\cuda{} block of our \nvidia{} A100 GPU is about $100$KB, this does not permit a multi-precision size of $2^{19}$ bits. 

Another limiting factor is the amount of register memory: currently \cuda{} supports a maximum of 255 registers per thread or $64$K registers per \cuda{} block, whichever is lowest. When the register demand exceeds these bounds, the \texttt{nvcc} compiler resorts to register spilling~\cite{cuda2014}, which allocates the excess registers in a higher level cache that is, however, significantly less efficient to access. 

Our implementation uses a maximal thread-sequentialization factor $Q=4$, since the implementation of multiplication~\cite{midsize-add-mul} is optimized for this value---i.e., each thread computes four elements of the multiplication result. Using $Q=4$ for the biggest multi-precision size ($2^{18}$ bits) results in spilling $40$ registers ($160$ bytes) to slower storage, each of them being accessed just under three times ($392$ bytes of spilled storage are accessed). In principle, suitably increasing $Q$ eliminates the spilling of registers---because it decreases the number of threads in a \cuda{} block and allows each thread to use more registers---but this did not improve the overall performance of the division implementation.   


\input{incl/fig-cuda-overview}

\subsection{Overview of the \cuda{} Implementation}

\Cref{fig:cuda-overview} gives the gist of the \cuda{} implementation by presenting several parallel components of the division algorithm. \Cref{lst:cpy} demonstrates the manner in which the input multi-precision integers are efficiently loaded from global to register memory using shared memory as a staging buffer: The first step uses consecutive threads to copy consecutive memory locations to shared memory (\texttt{shmem}), thus achieving coalesced reads from global memory. Each thread performs $Q$ such copies in the loop at lines~\ref{line:cpy-loop1-beg}-\ref{line:cpy-loop1-end}, where the \cuda{} block size is dimensioned such that \lstinline{M = Q * blockDim.x}. Since shared-memory accesses are not affected by un-coalesced accesses, the loop at lines~\ref{line:cpy-loop2-beg}-\ref{line:cpy-loop2-end} copies $Q$ consecutive elements from shared memory to each of the thread's private (register) memory.

\Cref{lst:shift} presents the operation that shifts a multi-precision integer $U$ by $n$: the result of the shift is manifested in shared memory by the first loop, and once all threads have terminated work (i.e., reached the \lstinline{__syncthreads()} barrier), the second loop loads $Q$ consecutive elements of the result to the register (private) memory of each thread. Please note that the shared-memory buffer is only transiently utilized, and the same buffer is reused for following operations.

Several operations of the whole-shifted inverse algorithm use arguments that are powers of the corresponding base $B$, e.g., subtracting or comparing with $B^{bpow}$. Such specialized cases accept a more efficient implementation than when the arguments are arbitrary integers.  \Cref{lst:specialized_sub} demonstrates the case of subtraction $U - B^{bpow}$: The first loop finds the lowest index $n$ of a non-zero digit of $U$ that is greater than or equal to $bpow$ across the $Q$ elements of each thread, and the \lstinline{atomicMin} operation uses hardware-supported atomics to efficiently extend this computation of $n$ across all threads. The second loop finalizes the implementation by subtracting one from each digit whose index is between $bpow$ and~$n$. This specialization significantly reduces the inter-thread communication required by the general implementation of subtraction, presented in the next section~\ref{subsec:sub}.

\Cref{lst:less-than} implements the general case of the less-than operator $U < V$ on arbitrary multi-precision integers $U$ and $V$.  Essentially, the loop computes the less than and equal to relations among the $Q$ digits of each thread. This is encoded in an \texttt{int} rather than a tuple of booleans: the first (least-significant) bit being set denotes less than (\texttt{U[i] < V[i]}), and the second bit being set denotes equality (\texttt{U[i]==V[i]}). The computation is extended across digits by means of the \texttt{LTop} operator (line~\ref{line:lt-apply-seq}) whose semantics on tuple-of-boolean arguments is:

\lstinline{def LTop (lt1, eq1) (lt2, eq2) = (lt2 || (eq2 && lt1), eq1 && eq2)}

\noindent where \texttt{lt1/2} denote whether the first/second digits are in a less-than relation and \texttt{eq1/2} similarly treats equality. The operator implements comparison across two digits: The less-than relation holds either when the second digits are in a less-than relation (\texttt{lt2}) or when the second digits are equal and the first digits conform with the less-than relation. Equality holds if both digits are equal. 

Finally, since \texttt{LTop} is associative, the computation is extended across all digits by performing a reduce with the \texttt{LTop} operator across the partial result of all threads. This is implemented as a prefix-sum (\texttt{scanBlk} at line~\ref{line:lt-scan}) followed by selecting the last element of the shared-memory result at line~\ref{line:lt-reduce}. Next section presents the \cuda{} implementation of prefix sum and the manner in which it is applied to implement subtraction.


\subsection{Subtraction of Multi-Precision Integers}
\label{subsec:sub}

We start by recounting the semantics of the map and exclusive scan (prefix sum) parallel skeletons. Scans are a classical primitive for parallel prefix computations~\cite{DBLP:conf/icpp/Blelloch87}:
map applies its function argument to each corresponding element of its input
arrays and exclusive scan computes all prefixes under an associative
operator $\odot$ having neutral element $e_\odot$:

\[
\mathrm{map} \ f \ [a_0,\ldots,a_{m-1}] \ = \ [f~ a_0,\ldots, f ~a_{m-1}]
\]
\[
\mathrm{map2} \ f \ [a_0,\ldots,a_{m-1}] \ [b_0,\ldots,b_{m-1}] \ = \ [f~ a_0~b_0,\ldots, f ~a_{m-1}~b_{m-1}]
\]
\[
    \mathrm{scan}^{\mathword{exc}} \ \odot \ e_\odot \ [a_0,\ldots,a_{m-1}] \ \equiv \ [e_\odot, \ a_0, \ a_0 \odot a_1, \ \ldots, \ a_0 \odot \ldots \odot a_{m-2}]
\]

\input{incl/fig-lst-sub-lst-scan}

Subtraction follows a similar procedure as the one used for addition in~\cite{midsize-add-mul}.  Assuming two multi-precision unsigned integers $x > y$, each having $m$ digits: $x = x_0\cdot B^0 + \ldots + x_{m-1}\cdot B^{m-1}$ and $y = y_0\cdot B^{0} + \ldots + y_{m-1}\cdot B^{m-1}$, their subtraction $x - y$ is computed by means of a map-scan-map composition:
\input{incl/lst-futhark-6}
where the \lstinline{|>} operator pipes the result of the preceding computation as the last argument of the following computation. Essentially, $\kw{map2}~ \ominus_{1} ~x ~y$ computes whether the (independent) per-digit subtraction results in underflow (\lstinline{a-b > a}) or in \lstinline{0}. The result is passed to $ \kw{scan}^{exc} ~\odot~ (\kw{false},\kw{true})$, which computes and propagates the carry for each digit, and finally, the last map performs the per-digit subtraction to which it applies the carry. The mapped operators $\ominus_{1}$ and $\ominus_3$ differ from the ones used for addition, but the combine operator $\odot$ is~the~same. 

Listings~\ref{lst:scan}~and~\ref{lst:sub} show the \cuda{} implementation: \cref{lst:scan} shows the implementation of the inclusive-scan operator at warp and block level, respectively, for a generic associative operator that accepts arguments of type \text{int}. For efficiency, the warp level uses register shuffling to avoid accessing shared memory.

\Cref{lst:sub} shows the implementation for subtracting \texttt{as - bs}, where inputs \texttt{as} and \texttt{bs} and the result \texttt{rs} are maintained in register memory and \texttt{as} is assumed larger than \texttt{bs}. The combine operator $\odot$, represented by class \texttt{CarryOP}, is optimized as in~\cite{midsize-add-mul} to encode the two boolean values as the least-significant two bits of a $32$-bit integer. Each thread performs the scan sequentially across its $Q$ elements (the loop at lines~\ref{line:sub-loop1-beg}-\ref{line:sub-loop1-end}) and publishes the final carry in shared memory (line~\ref{line:sub-thd-res}). A block level scan (\texttt{scanBlk}) computes and propagates the carries (line~\ref{line:sub-scan}), and the subtraction result is updated accordingly at lines~\ref{line:sub-update-beg}-\ref{line:sub-update-end}. 
For very small numbers of digits, hardware-inspired carry-lookahead or
carry-select schemes may have lower constant overhead, but our target
range is larger multi-precision integers fitting within a CUDA block;
accordingly, the implementation uses sequential propagation within each
thread and a block-level scan only across the per-thread summaries.


\newcommand{\hsp}{\hspace{3ex}}

\subsection{Implementation of Multi-Precision Multiplication}

\enlargethispage{\baselineskip}

The shift-inverse Algorithm~\ref{alg:shinv_revised} uses operations of different (increasing) sizes inside the loop in the \texttt{Refine} function. The operations that have cheap (linear) cost---e.g., comparison, subtraction---are implemented for simplicity by padding to the multi-precision size of the input. However, the multiplications have quadratic cost and need to be adjusted to the actual size of the arguments in order to allow the whole computation to have the same asymptotic as one full multiplication.

\input{incl/fig-full-conv}

Our implementation of size-variant multiplication builds on the one proposed in~\cite{midsize-add-mul} that addresses the fixed multi-precision case. \Cref{fig:full_conv} visually depicts the scheduling proposed in~\cite{midsize-add-mul} that minimizes inter-thread communication and ensures a load-balanced execution. Denoting by $m$ the multi-precision size and by $q$ the sequentialization factor and assuming $m$ and $q$ powers of two, the schedule assigns to each thread the computation of $q$ digits from the first half of the result and another $q$ symmetrical-opposite digits from the second half. It follows that each thread performs a fixed number $\frac{Q}{2} \cdot m$ of base (digit) multiplications.  \Cref{fig:full_conv} shows the cases $q=2$ and $q=4$ on the left- and right-hand sides. 

\enlargethispage{\baselineskip}

Our approach dispatches at runtime variant-sized multiplications to a number of statically-specialized instantiations of the multi-precision size:

\input{incl/lst-dynamic-mult}

The pseudocode above assumes that input \texttt{Ash} and \texttt{Bsh} and the result \texttt{Csh} are allocated in shared memory (where the result \texttt{Csh} overlaps with an input) and are relocated to registers inside the specializations. A small multiplication \texttt{smallMul} refers to the case when its precision is smaller than or equal to the \cuda{} block size \texttt{BLOCK}. This is essentially computed by allocating each digit of the result to a different thread or possibly by a procedure that maximizes thread utilization by using atomic accumulations in shared memory. The other cases call \texttt{effMul} that uses the efficient scheduling depicted in \Cref{fig:full_conv}: the second and third template arguments denote the \cuda{} block size and the sequentialization factor $q$; it follows that the result is computed in precision $q\cdot \texttt{BLOCK} \geq m$.  In practice we use $q=4$ as the maximal sequentialization factor.


We recognize that a polished library would not use just one multiplication method throughout --- that at different sizes one would transition between classical multiplication, Karatsuba or Toom-Cook multiplication  and a number theoretic transform (NTT/FFT) method.
For this work, however, we are content to study division in terms of whatever multiplication is used.

\section{Empirical Evaluation}
\label{sec:eval}


\paragraph{Hardware.} The evaluation was run under the Red Hat Enterprise Linux 8.10 operating system on an \nvidia{} A100 GPU, which has 6912 cores, a peak global memory bandwidth of 1,555 GB/sec and \texttt{FP32} peak performance of 19.5 TFLOP/s.
\vspace{-1ex}

\paragraph{Methodology.} The benchmarking setup is available at:
\smallskip

\url{https://github.com/aske0778/midint-arithmetic-division}  
\smallskip

\noindent and extends the framework presented in~\cite{midsize-add-mul} with the implementation of division $u/v$. Each problem instance is initialized with randomly generated integers. The precision of $u$ is fixed to $M-2$, such as to account for the two guard digits in the \texttt{Refine} function of Algorithm~\ref{alg:shinv_revised}. The precision of the divisor $v$ is randomly selected between $2$ and $M/2$. This configuration ensures that the refinement loop (in \texttt{Refine}) always performs the maximum number of iterations. 

The evaluation instantiates the digit type ($uint$) to $64$-bit unsigned integer and the (maximal) sequentialization factor to $Q=4$. The results are averaged across $25$ runs to minimize statistical variance. We compare the performance of our implementation with that of the \cgbn{} library~\cite{nvidialab:coopbignum} (namely \texttt{cgbn\_div\_rem}), which is also averaged over $25$ runs. \cgbn{} does not allow customization of the digits' word size or of the sequentialization factor.

\input{incl/tab-perf}

\paragraph{Datasets.} The datasets are chosen to vary the integer precision in number of bits (\texttt{Num Bits}) from $2^{13}$ to $2^{18}$ and the number of (parallel) division instances (\texttt{Num Insts}) from  $2^{19}$ down to $2^{14}$, such that $\texttt{Num Bits} \cdot \texttt{Num Insts} = 2^{32}$.

\enlargethispage{\baselineskip}

\paragraph{Validation.}
The results of our \cuda{} implementation match those of the GNU
Multi-Precision Library (\gmp) on all considered datasets. Beyond the
performance datasets, correctness testing used $1000$ randomized instances
with random values and effective precisions, plus manual tests of known
edge cases, including the shifted-inverse over-approximation that motivated
Theorem~\ref{thm:quo_by_shinv_opt}. The tests also covered all
multiplication sizes. All results agreed with \gmp{}, providing strong
evidence for correctness.

\paragraph{Performance Comparison with \cgbn.}
\Cref{tab:perf} compares our multiplication and division with \cgbn{}.
For multiplication, our implementation is faster on the two
highest-precision datasets by factors of about $35\times$ and $4.5\times$,
and by $1.21\times$ and $1.16\times$ on the next two; it breaks even on
precision $2^{14}$ and is slower by $1.08\times$ on the last one. 
This is consistent with the different design points: \cgbn{} uses a warp-level
organization,
whereas our implementation targets larger precisions with one arithmetic
instance per \cuda{} block. A precise attribution of the gap to register
pressure, scheduling, memory transactions, or occupancy would require a
separate counter-level study, which is outside the scope of this paper.

The fifth column shows the factor by which our implementation of division is slower than one full multiplication that is used in its implementation. The highest-precision datasets (of precisions $2^{18}$, $2^{17}$, $2^{16}$) offer good performance, being (only) $6.77-6.94\times$ slower than a full multiplication---
recall that Section~\ref{subsec:cost} argued that the target division algorithm requires at least $5$ full multiplications.

\enlargethispage{\baselineskip}

%
%
The last three datasets exhibit significant loss of performance as high as $3\times$ away from the optimal, i.e., from about $10-15\times$ slower than one full multiplication. The performance loss is due to:
\begin{itemize}
\item assigning a suboptimal number of threads per \cuda{} block---the current implementation computes one division instance with a \cuda{} block of threads, which, for example, has suboptimal size $32$ for the last dataset. 
\item the fact that the weight of the full multiplications in the total runtime decreases with the integer precision---the other operations become significant.   
\end{itemize}

The sixth column compares the performance of our division with that of \cgbn{}. Importantly, \cgbn{} does not support division at precisions higher than $2^{15}$ bits, although multiplication is supported suboptimally up to $2^{18}$ bits. On the last three lower-precision datasets, \cgbn{} division is faster by significant factors ranging from $3.6\times$ to $7.4\times$. In particular \cgbn's division is only around $2.5\times$ slower than one \cgbn{} full multiplication. 
This indicates that, for the lower precision range up to $2^{15}$ bits,
the \cgbn{} implementation is better matched to the problem size than the
block-level division method studied in this paper,\footnote{
  This is also why we did not optimize our implementation of division for the last three datasets---even reaching the optimal cost of five full multiplications would still not close the performance gap with \cgbn.
}
while the block-level method covers precisions higher than $2^{15}$ and up to at least $2^{18}$ bits. The last column shows the speedup of our division {\em vs.} GMP.

The present comparison is therefore intended to test the multiplication-based
cost model and to identify the precision range where the shifted-inverse
method is promising, rather than to provide a complete production-library
benchmark.  Further engineering work could combine a counter-level analysis
with refinements such as clipped products~\cite{clipped-prods}, which could
reduce the cost of the double-sized multiplication used in line~\ref{line:quot-init}
of Algorithm~\ref{alg:div_new}.

\section{Related Work}
\label{sec:rel-work}

This paper reports a \cuda{} implementation of the multi-precision division algorithm proposed in~\cite{watt2023}, which extends the Newton method to be carried out entirely in the integral domain by means of data-parallel operations such as shifting and multiplying multi-precision integers. This algorithm can in principle be further improved with the refinement of clipped products~\cite{clipped-prods}, which would enable cheaper computation of the middle digits of a multiplication result.

The implementation presented in this paper uses the \cuda{} implementations for addition and multiplication reported in~\cite{midsize-add-mul}, which also inspires the parallelization strategy used for division. The work also reports matching implementations in the high-level data-parallel Futhark language. The Futhark compiler provides a set of useful optimizations, e.g., related to autotuning the degree of exploited parallelism~\cite{futhark-tuning}, mapping computations to \cuda{} block level with allocation of intermediate results in shared memory~\cite{futhark-ppopp}, reusing memory buffers~\cite{futhark-mem-sc22}, automatic differentiation~\cite{ifl-rev-ad,futhark-ad-sc22}, static~\cite{pldi-idx-arr-prop} or dynamic~\cite{henriksen2014bounds} verification of memory safety. These have enabled efficient acceleration of applications from various domains~\cite{bfast,annf,STL-rem-sens}, which were exported as python libraries~\cite{futhark-python}. However, Futhark's multiplication suffers significant loss of performance because Futhark lacks code transformations aimed at mapping parallel arrays in register rather than shared memory.  We have implemented the target division algorithm in Futhark as well;  its performance is poor because the mentioned shortcomings are exacerbated by the nature of the division algorithm that requires parallel operations whose sizes vary through a loop, e.g., shifts and multiplications.

The closest related work to the one in this paper is the “Cooperative Groups Big
Numbers” (\cgbn) library~\cite{nvidialab:coopbignum}, authored by NvidiaLab, that offers high-performance implementation for multi-precision integers up to $32$K bits, including multiplication and division. The key technique enabling high performance in \cgbn{} is to map an instance of integer computation on at most one warp of threads in order to leverage specialized \nvidia{} hardware that allows values to be communicated directly between registers (of the warp), i.e., avoiding the overhead of accessing the shared memory, which has significantly higher latency.  However, \cgbn{} does not support division on precisions higher than $32$K bits.
%
%
Other less related works targeting parallel multi-precision arithmetic include:
\begin{itemize}
    \item Previous work has investigated the feasibility of retargeting the multi-precision algorithms used by \gmp{} for GPU execution. The conclusion has been that the architectural differences do not allow easy and/or efficient porting, which motivated development of new, inherently parallel algorithms~\cite{emmart2018}.

    \item Efficient parallel algorithms for division have been devised to cover specialized input~\cite{emmart2011}. For example, the case where the divisor is a digit has been efficiently solved with a modified version of Jebelean's exact division algorithm, which has parallel complexity $O(n/p + \log p)$, where $n$ denotes the precision and $p$ the number of processors.

    \item CAMPARY~\cite{Campary1,Campary2} supports sequential CPU and GPU execution of multi-precision floating-point arithmetic up to a few hundred bits, e.g., addition, multiplication, division, square root. 
    The key idea is to represent real numbers as unevaluated sums of multiple machine-precision floating point values.

    \item Older libraries aimed at supporting the functionality of \gmp{} in \cuda{} include GARPREC~\cite{GARPREC} and CUMP~\cite{Cump1,Cump2}. The latter offers precision up to about $62$ decimal digits and its objective was to outperform GARPREC.  Similar works were ancestors of \cgbn, with which~we~compare~directly.

    \item Isupov~\cite{Isupov-GRNS} uses interval techniques to augment residue number arithmetic for operations that rely on magnitude for numbers with up to 4096 bits.

    \item A rich body of work~\cite{cuFFT-base10,Moreno-GPU-FFT,JHPCA-FFT-10bits,ParProcLetters-AddSubMul-early2010} was aimed at accelerating multi-precision integer multiplication on GPUs using adaptations of the classical quadratic algorithm and of Strassen's log-linear algorithm~\cite{strassen1971schnelle}.
\end{itemize}

Libraries supporting multi-precision arithmetic expose highly-parameterized components (tower of generic types). For example, the division algorithm is parameterized over the underlying multiplication algorithm---e.g., classical, Karatsuba or Strassen---which in turn imposes different representations for the multi-precision integers. Interoperating such libraries with various DSLs and mainstream computer-algebra systems~\cite{Maple,Mathematica}, which support different mechanisms for parametrization, remains a direction of interest~\cite{mapal_synasc,alma:ISSAC,gidl_oopsla}.

We note that algorithms for multi-precision integer arithmetic are closely related to algorithms for univariate polynomials with coefficients in $\mathbb{Z} \mod m$.  Of particular interest are those with $m$ a word-sized prime, for which there are good GPU algorithms.  An important complication with multi-precision integers is the need to handle carries and borrows.

\section{Future Work and Conclusions}
\label{sec:conclusion}

This paper has examined multi-precision integer division on GPUs in a
midsize precision range: larger than the range where warp-level
cooperative libraries are most effective, but still small enough that a
complete division instance can be assigned to a single \cuda{} block.  In
this regime, the whole-shifted-inverse algorithm of
Watt~\cite{watt2023} is a good structural match for GPU execution.  Its
Newton refinement is expressed using integer multiplication, shifting,
comparison, subtraction, and related data-parallel operations, avoiding
the need to move into a floating-point reciprocal domain.

The implementation maps one division instance to a \cuda{} block, keeps
active operands in registers where possible, uses shared memory for
communication and staging, and dispatches variable-size multiplications
to statically specialized implementations. These choices reduce global
memory traffic and expose the compiler transformations needed for a
high-level data-parallel language such as Futhark to generate similar
code.

The empirical results support the cost model developed in
Section~\ref{subsec:cost}.  For the largest tested precision, division is
only slightly more expensive than the predicted lower bound of five full
multiplications.  This indicates that, at these sizes, the overheads from
shifts, comparisons, subtractions, and control structure have been kept
small relative to the dominant multiplication costs.  At smaller
precisions these overheads become more visible, and \cgbn{} remains faster
where its division implementation is available.  Thus the present
implementation should be viewed not as a replacement for warp-level
libraries in their strongest range, but as evidence that a block-level
strategy can effectively cover larger midsize precisions.

Several directions remain open.  The most immediate algorithmic
improvement is to use clipped products~\cite{clipped-prods}.  In the
current implementation, the quotient approximation requires a
double-precision multiplication even though only a selected range of
product digits is ultimately needed.  A clipped-product implementation
could reduce this cost and bring the observed runtime closer to the
idealized multiplication count.

A second direction is better specialization across operand sizes.  The
current implementation uses a fixed block-level strategy and a maximal
sequentialization factor chosen to work well for the largest tested
inputs.  This is not optimal for smaller inputs, where the number of
active threads per division instance can be too low and the relative cost
of non-multiplication operations increases.  More aggressive autotuning
of block size, sequentialization factor, and multiplication
specialization would likely improve performance across the full tested
range.

A third direction is integration with high-level GPU programming systems.
The \cuda{} implementation identifies several transformations that are
important for performance: keeping short arrays in registers rather than
shared memory, staging global-memory accesses to preserve coalescing,
specializing operations by effective precision, and avoiding unnecessary
materialization of intermediate arrays.  Division is a useful stress test
for such compiler work because it combines scans, shifts, comparisons,
variable-size multiplications, and iterative refinement in a single exact
arithmetic computation.

Finally, larger precisions raise the question of asymptotically faster
multiplication.  The shifted-inverse division algorithm is parameterized
by multiplication, and can therefore in principle benefit from Karatsuba,
FFT, or NTT-based methods.  On GPUs, however, these methods introduce
their own memory-management constraints.  Near the fast-memory limit, CRT
or NTT-based products may require several modular products for the
largest multiplication, while shorter products arising during the Newton
refinement may be able to use fewer moduli or longer transform vectors.
This suggests that fast multiplication should not be treated as a
black-box replacement only; it should be integrated with the changing
precision requirements of the division algorithm.

In summary, the results show that exact multi-precision integer division
based on whole shifted inverses can be implemented efficiently on GPUs
for midsize operands.  The implementation reaches near-model performance
at the largest tested precisions, covers sizes beyond those supported by
\cgbn{} division in our experiments, and clarifies the low-level operations
and compiler transformations needed for future high-level
implementations.

\bibliographystyle{splncs04}
\bibliography{references.bib}

%
%
%
%
%
%
%
%
%

\end{document}

%% file: language.tex
\newcommand{\nvidia}[0]{\textsc{nvidia}}
\newcommand{\gmp}[0]{{\textsc{gmp}}}
\newcommand{\cuda}[0]{{\textsc{cuda}}}
\newcommand{\cgbn}[0]{{\textsc{cgbn}}}

\newcommand{\kw}[1]{\mbox{\texttt{\bfseries{#1}}}}



%% file: futhark.tex
\lstdefinelanguage{futhark}
{
  morekeywords={
    alloc,
    scan,
    map,
    map2,
    map3,
    map4,
    concat,
    def,
    do,
    else,
    f32,
    u32,
    bool,
    for,
    fun,
    if,
    in,
    include,
    let,
    loop,
    struct,
    then,
    type,
    val,
    while,
    with,
    mapnest,
    module,
    where,
    sort,
    scratch,
    multired,
    reverse,
    zip3,
    unzip3,
    copy,
    gather,
    withacc,
    sum
  },
  sensitive=true, 
  morecomment=[l]{--}, 
  morecomment=[s]{\{-}{-\}}, 
  morestring=[b]", 
  literate={\\}{\fn}{1} {->}{$\rightarrow$}{1} {<-}{$\leftarrow$}{1} {|>}{$\pipe$}{1},
}

\usepackage{xcolor}
\definecolor{eclipseBlue}{RGB}{42,0.0,255}
\definecolor{eclipseGreen}{RGB}{63,127,95}
\definecolor{eclipsePurple}{RGB}{127,0,85}

\newcommand{\pipe}{\ensuremath{\triangleright}}

\definecolor{comment}{RGB}{0,128,0} 
\definecolor{string}{RGB}{255,0,0}  
\definecolor{types}{RGB}{200,150,0}  
\definecolor{keyword}{RGB}{0,0,255} 

\lstdefinestyle{cuda}{
	commentstyle=\color{comment},
	stringstyle=\color{string},
	keywordstyle=\color{keyword},
	basicstyle=\footnotesize\ttfamily,
	numbers=left,
	numberstyle=\tiny,
	numbersep=5pt,
	frame=lines,
	breaklines=true,
	prebreak=\raisebox{0ex}[0ex][0ex]{\ensuremath{\hookleftarrow}},
	showstringspaces=false,
	upquote=true,
	tabsize=2,
    morekeywords={OP, __syncthreads, blockDim, threadIdx, blockIdx},
    captionpos=b,
}

%% file: incl/alg-shinv-revised.tex
\begin{algorithm}[tp]
\label{alg:shinv_revised}
\DontPrintSemicolon
\LinesNumbered
\caption{\textsc{Shinv}($v$, $h$, $B$) in $\mathbb{Z}$}
\KwIn{$v, \;h,\; B \in \mathbb{Z}_{>0},\; B^k \leq v < B^{k+1}$}
\KwOut{$\textsc{Shinv}_{h,B}(v) = \left\lfloor \frac{B^h}{v} \right\rfloor$ \ \ {\color{gray} $\triangleright$ All shifts are w.r.t. $B$}}
\KwUse{\textsc{Mult}, a multi-precision multiplication method\\
    \hspace*{3em}\textsc{PowDiff}, to compute $B^h - v \cdot w$ (Algorithm~\ref{alg-powdiff})\\
    \hspace*{3em}\textsc{shift}, a shift operation for multi-precision ints}
    \medskip

\SetKwFunction{FShinv}{Shinv}
\SetKwProg{Fn}{Function}{:}{}
\Fn{\FShinv{$v$, $h$, $B$}}{
    \textcolor{black}{\color{gray} $\triangleright$ Group digits if base is small}\; \label{g1-start}
    \If{$B < 16$}{
    $p \gets \min(6-B, 2)$\;
    \KwRet $\textsc{shift}_{h \rem p-p} ~(~\textsc{Shinv}(v,h \quo p+1,B^p)~)$
    }
    \textcolor{black}{\color{gray} $\triangleright$ Special cases guarantee $B < v \leq B^h / 2$}\;
    \lIf{$v < B$}{\KwRet $B^h \quo  v$ \qquad {\color{gray} $\triangleright$ Divide by 1 digit} }
    \lIf{$v > B^h$}{\KwRet $0$}
    \lIf{$2v > B^h$}{\KwRet $1$}
    \lIf{$v = B^k$}{\KwRet $B^{h-k}$}  \label{g1-end}
    \textcolor{black}{\color{gray} $\triangleright$ Form initial approximation}\; \label{g2-start}
    \stephenNote{they deemed that returning the initial approx}
    \stephenNote{is incorrect; as well they simplified it to 2 digits (rather than 3)}
    $V \gets v_{k-1} + v_k\cdot B$\;
    $w \gets B^{3}  \quo V$ \qquad {\color{gray} $\triangleright$ Divide 4 digits by 2 digits}\; \label{g2-end}
    \Return $\textsc{Refine}(v, h, k, w, 2)$ \qquad {\color{gray} $\triangleright$ Refine $w$ iteratively}\;
}
\SetKwFunction{FRefine}{Refine}
\SetKwProg{Fn}{Function}{:}{}
\Fn{\FRefine{$v$, $h$, $k$, $w$, $\ell$}}{
    $g \gets 2$ \qquad {\color{gray} $\triangleright$ Guard digits}\;
    $w \gets \textsc{shift}_g(w)$ \;

\textcolor{black}{\color{gray} $\triangleright$ loops at least 2 iters; otherwise similar to while(l < h - k)}\;
\For{$i \gets 0$; $i < \left\lceil \max\left(\log_2(h-k-1),\; 0 \right) \right\rceil + 2$; $i\texttt{++}$} {
        $m \gets \min(h - k + 1 - \ell,\; \ell)$ \;
        $s \gets \max(0, \;k - 2\ell + 1 - g)$ \qquad {\color{gray} $\triangleright$ How to scale $v$}\;
        $w \gets \textsc{Step}(k + \ell + m - s + g, \; \textsc{shift}_{-s}(v), \;w, \;m, \; \ell, \; g)$ \;  

        \stephenNote{Then branch is new (their code also validates)}
        
        \lIf{$i<2$}{$w \gets \textsc{shift}_{-m}(w)$} 
        \Else{ 
        $w \gets \textsc{shift}_{-1}(w)$\;
        $\ell \gets \ell + m - 1$
        }
    }
    \stephenNote{they shift with q rather than g as computed below:}
    $q \gets (h-k < 2) ~\text{?}~ h-k-4 ~\text{:}~ -2$ \;
    \Return $\textsc{shift}_{q}(w)$
}
\SetKwFunction{FStep}{Step}
\SetKwProg{Fn}{Function}{:}{}
\Fn{\FStep{$h$, $v$, $w$, $m$, $\ell$, $g$}}{
    $(sign, \; x) \gets \textsc{PowDiff}(v, w, h - m, \ell - g, B)$\;
    \stephenNote{then branch is as original; else is new. Can PowDiff return a negative?}
    \lIf{$sign$} {\KwRet  {$\textsc{shift}_m(w) + \textsc{shift}_{2m - h}(\textsc{Mult}(w, x))$} }
    \Else{
    $tmp \gets \textsc{Mult}(w, x)$\;
    $res \gets \textsc{shift}_m(w) - \textsc{shift}_{2m - h}(tmp)$\;
    \If{any of the $2m-h$ least significant digits of $tmp$ are nonzero}{
    $res \gets res -1$\;
    }
    \Return $res$}
}
\end{algorithm}

%% file: incl/alg-powdiff.tex
\begin{algorithm}[tp]
\label{alg-powdiff}
\DontPrintSemicolon
\caption{PowDiff($v$, $w$, $h$, $\ell$, $B$) in $\mathbb{Z}$}
\KwIn{$v, \;w,\; h,\; \ell,\; B \in \mathbb{Z}_{>0}$ such that $\text{prec}|w - \textsc{shinv}_hv| \leq \text{prec}(w) - \ell$}
\KwOut{$(sign, \; |B^h - v \cdot w|)$}
\KwUse{ \textsc{Mult}($a$, $b$) = $a \cdot b$\\
    \hspace*{3.2em}\textsc{MultMod}($a$, $b$, $d$, $B$) = $(a \cdot b) \bmod B^d$}
    
\SetKwFunction{FPowDiff}{PowDiff}
\SetKwProg{Fn}{Function}{:}{}
\Fn{\FPowDiff{$v$, $w$, $h$, $\ell$, $B$}}{
    $L \gets \text{prec}_Bv + \text{prec}_Bw - \ell + 1$\;
    \stephenNote{the original code was modified to explicitly track the sign since we do not support negative integers. Otherwise looks identical.}
    
    \If{$v = 0 \vee w = 0 \vee L \geq h$}{
        \lIf{$B^h > \textsc{Mult}(v, w)$}{\KwRet $(1, \; B^h - \textsc{Mult}(v, w))$}
        \lElse{\KwRet $(0, \; \textsc{Mult}(v, w) - B^h)$}}
    \Else{
        $P \gets \textsc{MultMod}(v, w, L, B)$\;
        \lIf{$P = 0$}{\Return $(1,\;0)$}
        \lElseIf{$P_{L-1} = 0$}{\Return $(0, \;P)$}
        \lElse{\Return $(1,\; B^L - P)$}
    }
}
\end{algorithm}

%% file: incl/alg-div-new.tex
\begin{algorithm}[tp]
\label{alg:div_new}
\DontPrintSemicolon
\LinesNumbered
\caption{Div($u$, $v$, $m$, $B$)} 
\KwIn{$u, \;v \in \mathbb{Z}_+^m,\; m,\; B \in \mathbb{Z}_+$}
\KwOut{$(q, r) \ \text{with} \ u = q\cdot v + r,~r < v$}
\KwUse{ \textsc{Mult}($a$, $b$) = $a \cdot b$\\
        \hspace*{3.2em}\textsc{Prec}, to compute precision \\
       
       \hspace*{3.2em}\textsc{Shift}, for shifting the integer}
\medskip
    
    $h \gets $prec$(u)$ \hspace{13ex} \textcolor{black}{\color{gray} $\triangleright$ precision of $u$}

    \stephenNote{can we avoid the double }
    \stephenNote{multiplication below and one for r?}
    
    $q \gets \shift_{-h}(~ \textsc{Mult}(~u, ~\shinv(v,h,B)~)~)$  {\color{gray} $\triangleright$ Initial Quotient} \label{line:quot-init}

    $m \gets \textsc{Mult}(v, ~q)$ \label{line:rem-init}
    
    \If{$u < m$}{
        $q \gets q - 1$  \hspace{9ex} \textcolor{black}{\color{gray} $\triangleright$ Handles $\delta = -1$}
        
        $m \gets m - v$
    }

    $r \gets u - m$   \hspace{15ex} {\color{gray} $\triangleright$ Initial Remainder} 
    
        
    \If{$r \geq v$}{
        $q \gets q + 1$  \hspace{11ex} \textcolor{black}{\color{gray} $\triangleright$ Handles $\delta = 1$}
        
        $r \gets r - v$
    }

    \Return{$(q, r)$}
\end{algorithm}

%% file: incl/fig-cuda-overview.tex
\begin{figure}[tp]
\begin{minipage}{0.48\textwidth}
\begin{lstlisting}[language=c++, style=cuda, label=lst:cpy, basicstyle=\ttfamily\scriptsize, escapechar=|, caption=Coalesced copy of integers from global (\texttt{AGlb}) to shared (\texttt{sh\_mem}) to register memory (\texttt{AReg}).]
template< class uint, 
          uint32_t M, uint32_t Q >
__device__ inline void 
cpyGlb2Reg( uint* AGlb
          , volatile uint* shmem
          , uint AReg[Q]
) {
  const uint32_t
    glb_off = blockIdx.x * M;

  for (int i = 0; i < Q; i++) { |\label{line:cpy-loop1-beg}|
    int idx = i * blockDim.x + 
              threadIdx.x;
    shmem[idx]= AGlb[idx+glb_off];
  }                             |\label{line:cpy-loop1-end}|
  __syncthreads();
  for (int i = 0; i < Q; i++)   |\label{line:cpy-loop2-beg}|
   AReg[i]=shmem[Q*threadIdx.x+i]; |\label{line:cpy-loop2-end}|
}
\end{lstlisting}
\end{minipage}
\begin{minipage}{0.49\textwidth}
\begin{lstlisting}[language=c++, style=cuda, label=lst:shift, numbers=none,  basicstyle=\ttfamily\scriptsize, caption=Shift integer \texttt{U} by \texttt{n}: input \& result \texttt{R} are held in registers; shared memory (\texttt{sh\_mem}) is used as staging buffer.]
template< class uint, 
          uint32_t M, uint32_t Q >
__device__ inline void 
shift( int n, uint U[Q]
     , volatile uint* shmem
     , uint R[Q]               ) {
  for (int i = 0; i < Q; i++) {
    int idx = Q * threadIdx.x + i;
    int offset = idx + n;
    uint val = 0;
    if (offset >= 0 && offset < M)
         val = U[i];
    else offset = M-idx-1;
    shmem[offset] = val;
  }
  __syncthreads();
  for (int i = 0; i < Q; i++)
    R[i] = shmem[Q*threadIdx.x+i];
}
\end{lstlisting}
\end{minipage}\\
\begin{minipage}{0.48\textwidth}
\begin{lstlisting}[language=c++, style=cuda, label=lst:specialized_sub, numbers=none, basicstyle=\ttfamily\scriptsize, caption=Subtraction \texttt{U - $B^{bpow}$}]
template<class uint, uint32_t Q>
__device__ inline void 
subPowB( uint U[Q], uint32_t bpow
       , volatile uint* shmem  ) {
  uint32_t n = UINT32_MAX;
  if(threadIdx.x==0) shmem[0] = n;
  __syncthreads();
  // find lowest non-zero digit whose index n >= bpow
  for (int i = 0; i < Q; i++) {
    int rev_i = Q - i - 1;
    int idx= Q*threadIdx.x + rev_i;
    if (U[rev_i]!=0 && idx>=bpow)
      n = idx;
  }
  atomicMin((uint32_t*)shmem, n);
  __syncthreads();
  // subtract one from all digits between bpow and n
  uint32_t nn = shmem[0];
  for (int i = 0; i < Q; i++) {
    uint32_t idx = Q*threadIdx.x+i;
    if (idx >= bpow && idx <= nn)
      U[i] = U[i] - 1;
} }
\end{lstlisting}
\end{minipage}
\begin{minipage}{0.51\textwidth}
\begin{lstlisting}[language=c++, style=cuda, label=lst:less-than, escapechar=@, basicstyle=\ttfamily\scriptsize, caption=Less-than operator: \texttt{U<V}]
class LTop { public:
  static __device__ inline 
  int apply(int a, int b) {
   int r = a & b & 2;
   r +=(b & 1) || ((b & 2)&&(a & 1));
   return r;
  } 
};
template<class uint, uint32_t Q>
__device__ inline bool 
lt( uint U[Q], uint V[Q]
  , volatile uint* shmem ) {
  int R[Q] = {0};
  for (int i = 0; i < Q; i++) { @\label{line:loop-beg}@
    if (U[i] < V[i])            @\label{line:lt-1}@
      R[i] |= 1;     
    else if (U[i] == V[i])      @\label{line:lt-2}@
      R[i] |= 2;
    if (i > 0)
      R[i] = LTop::apply        @\label{line:lt-apply-seq}@
               (R[i-1], R[i]);
  }                             @\label{line:loop-end}@
  scanBlk<LTop>(R[Q-1],shmem);  @\label{line:lt-scan}@
  return shmem[blockDim.x-1] & 1; @\label{line:lt-reduce}@
}\end{lstlisting}
\spacetune{\vspace{2pt}}
\end{minipage}
\caption{\cuda{} code for (1) copying in coalesced way from global to register memory, (2) shifting multi-precision integers, (3) subtracting a power of $B$ and for (4) lower-than comparison of multi-precision integers.}
\label{fig:cuda-overview}
\end{figure}

%% file: incl/fig-lst-sub-lst-scan.tex
\begin{figure}[tp]
\begin{minipage}{0.53\textwidth}
\begin{lstlisting}[language=c++, style=cuda, escapechar=@, label=lst:sub, basicstyle=\ttfamily\scriptsize, caption=Subtracting multi-precision integers: \texttt{as - bs} where \texttt{as > bs}.]
class CarryOP { public: ...
  static int identity() { return 2;}
  static int apply(int c1, int c2) {
   return ( c1 & c2 & 2 ) |
    ((( c1 & ( c2 >> 1)) | c2) & 1);
} };
template<class D,class S,uint32_t Q>
__device__  void
subRegs ( D as[Q], S bs[Q], D rs[Q]
        , volatile int* shmem ) {
  int cs[Q];
  int carry = CarryOP::identity();
  for(int i=0; i<Q; i++) {    @\label{line:sub-loop1-beg}@
    rs[i] = as[i] - bs[i];
    int c = rs[i] > a[i];
    c = c | ((rs[i] == 0) << 1);
    carry= CarryOP::apply(carry, c);
    cs[i] = c;
  }                 @\label{line:sub-loop1-end}@
  shmem[threadIdx.x] = carry;@\label{line:sub-thd-res}@
  __syncthreads(); 
  scanBlk<CarryOP>(shmem,threadIdx.x);@\label{line:sub-scan}@
  carry = CarryOP::identity();@\label{line:sub-update-beg}@
  if(threadIdx.x > 0)                   
    carry = shmem[threadIdx.x - 1];
  for(int i=0; i<Q; i++) {           
    rs[i] -= (carry & 1);
    carry=CarryOP::apply(carry,cs[i]);@\label{line:sub-update-end}@
} }
\end{lstlisting}
\end{minipage}
\begin{minipage}{0.46\textwidth}
\begin{lstlisting}[language=c++, style=cuda, numbers=none, label=lst:scan, basicstyle=\ttfamily\scriptsize, caption=Inclusive scan at\\ warp and block levels.]
template<class OP> __device__ 
int scanWarp( int u, int lane) {
  for(int i=1; i < 32; i *= 2) {
    int go = (lane >=i) ? i : 0;
    int elm = __shfl_up_sync
            (0xFFFFFFFF, u, go);
    u = OP::apply(elm, u);
  }
  return u;
}
template<class OP> __device__ 
int scanBlk(uint32_t u
        , volatile int* shmem){
  int idx = threadIdx.x;
  const int lane = idx & 31, 
            warpid = idx >> 5;
  int r = scanWarp<OP>(u, lane);
  if(lane==31) shmem[warpid] =r; 
  __syncthreads();
  if (warpid == 0)
   scanWarp<OP>(shmem[idx],lane);
  __syncthreads();
  if (warpid > 0)
  r=OP::apply(shmem[warpid-1],r);
  __syncthreads();
  shmem[idx] = r;
  __syncthreads();
  return r;
}
\end{lstlisting}
\end{minipage}
\end{figure}

%% file: incl/lst-futhark-6.tex
\begin{lstlisting}[language=futhark, mathescape=true, basicstyle=\ttfamily\small]
def $\ominus_{1}$ a b = (a - b > a, a - b == 0)
def $\odot$ (uf1, z1) (uf2, z2) = ( (uf1 && z2) || uf2, z1 && z2 ) 
def $\ominus_{3}$ a b (c, _) = a - b - c
def sub x y = map2 $\ominus_{1}$ x y |> scan$^{exc}$ $\odot$ (false,true) |> map3 $\ominus_{3}$ x y
\end{lstlisting}

%% file: incl/fig-full-conv.tex
\begin{figure}[tp]
\begin{minipage}{0.4\textwidth}
    \[ \begin{array}{clccc}
    \color{red}{c_0} & = \sum\limits_{i+j=0} a_i \cdot b_j & \hspace{1ex} & \mapsto & \color{red}{t_0}\\
    \color{blue}{c_1} & = \sum\limits_{i+j=1} a_i \cdot b_j & \hspace{1ex} & \mapsto & \color{blue}{t_1}\\
    \color{darkgreen}{c_2} & = \sum\limits_{i+j=2} a_i \cdot b_j & \hspace{1ex} & \mapsto & \color{darkgreen}{t_2}\\
    \color{cyan}{c_3} & = \sum\limits_{i+j=3} a_i \cdot b_j & \hspace{1ex} & \mapsto & \color{cyan}{t_3}\\
    \vdots & \hfill \vdots \hfill & & \hfill & \vdots \hfill \\
    \color{cyan}{c_{m-4}} & = \sum\limits_{i+j=m-4} a_i \cdot b_j & \hspace{1ex} & \mapsto & \color{cyan}{t_3}\\
    \color{darkgreen}{c_{m-3}} & = \sum\limits_{i+j=m-3} a_i \cdot b_j & \hspace{1ex} & \mapsto & \color{darkgreen}{t_2}\\
    \color{blue}{c_{m-2}} & = \sum\limits_{i+j=m-2} a_i \cdot b_j & \hspace{1ex} & \mapsto & \color{blue}{t_1}\\
    \color{red}{c_{m-1}} & = \sum\limits_{i+j=m-1} a_i \cdot b_j & \hspace{1ex} & \mapsto & \color{red}{t_0}\\
    \end{array} \]
\end{minipage}
\begin{minipage}{0.58\textwidth}
    \[ \begin{array}{clccccr}
    \color{red}{c_0} & = \sum\limits_{i+j=0} a_i \cdot b_j & \hspace{1ex} & \mapsto & \color{red}{t_0} & \hsp & 1~\text{term}\\
    \color{red}{c_1} & = \sum\limits_{i+j=1} a_i \cdot b_j & \hspace{1ex} & \mapsto & \color{red}{t_0} & \hsp & 2~\text{terms}\\
    \color{blue}{c_2} & = \sum\limits_{i+j=2} a_i \cdot b_j & \hspace{1ex}  & \mapsto & \color{blue}{t_1} & \hsp & 3~\text{terms}\\
    \color{blue}{c_3} & = \sum\limits_{i+j=3} a_i \cdot b_j & \hspace{1ex}  & \mapsto & \color{blue}{t_1} & \hsp & 4~\text{terms}\\
    \vdots & \hfill \vdots \hfill & & \hfill & \vdots \hfill & & \\
    \color{blue}{c_{m-4}} & = \sum\limits_{i+j=m-4} a_i \cdot b_j & \hspace{1ex}  & \mapsto & \color{blue}{t_1} & \hsp & m-4~\text{terms}\\
    \color{blue}{c_{m-3}} & = \sum\limits_{i+j=m-3} a_i \cdot b_j & \hspace{1ex} & \mapsto & \color{blue}{t_1} & \hsp & m-3~\text{terms} \\
    \color{red}{c_{m-2}} & = \sum\limits_{i+j=m-2} a_i \cdot b_j & \hspace{1ex}  & \mapsto & \color{red}{t_0} & \hsp & m-2~\text{terms}\\
    \color{red}{c_{m-1}} & = \sum\limits_{i+j=m-1} a_i \cdot b_j & \hspace{1ex}  & \mapsto & \color{red}{t_0} & \hsp & m-1~\text{terms}
    \end{array} \]
\end{minipage}
    \caption{Embarrassingly-parallel load-balanced scheduling of quadratic/classical multiplication proposed by \cite{midsize-add-mul}: The left schedule uses $\frac{m}{2}$ threads each computing $q=2$ elements of the result by performing $m$ digit multiplications. The one on the right uses $m/4$ threads, each computing $q=4$ elements by performing $2\cdot m$ digit multiplications.   
    }
    \label{fig:full_conv}
\end{figure}

%% file: incl/lst-dynamic-mult.tex
\begin{lstlisting}[language=c++, style=cuda, numbers=none, label=lst:dynamic_mult]
if (m <= BLOCK) {
    smallMul<uint>(m, Ash, Bsh, Csh); 
} else if (m <= 2 * BLOCK) {
    effMul<uint, BLOCK, 2>(Ash, Bsh, Csh);
} else if (m <= 4 * BLOCK) {
    effMul<uint, BLOCK, 4>(Ash, Bsh, Csh);
} else if (m <= 8 * BLOCK) ...
\end{lstlisting}

%% file: incl/tab-perf.tex
\begin{table}[tp]
\centering
\resizebox{0.7\textwidth}{!}{%
\begin{tabular}{|c|c||r|c||c|c|c|}
\hline
Num  & Num   & \textsc{Our Mul} & \textsc{Cgbn Mul} / & \textsc{Our Div} / & \textsc{Our Div} / & \textsc{Gmp Div} /\\
Bits & Insts & time (ms)        & \textsc{Our Mul}  & \textsc{Our Mul}   & \textsc{Cgbn Div} & \textsc{Our Div}
\\\hline\hline
$2^{18}$ & $2^{14}$ & 207.3 $ms$ &  35.1$\times$   & 6.77$\times$    &  \textcolor{red}{$-$} & 11.3$\times$\\\hline
$2^{17}$ & $2^{15}$ & 105.5 $ms$ &  4.48$\times$   & 6.66$\times$    &  \textcolor{red}{$-$} & 14.0$\times$\\\hline
$2^{16}$ & $2^{16}$ &  57.0 $ms$ &  1.21$\times$   & 6.94$\times$    &  \textcolor{red}{$-$} & 18.9$\times$\\\hline
$2^{15}$ & $2^{17}$ &  30.6 $ms$ &  1.16$\times$   & 9.73$\times$    &  3.63$\times$ & 12.6$\times$\\\hline
$2^{14}$ & $2^{18}$ &  19.3 $ms$ &  1.01$\times$   & 14.74$\times$    &  4.98$\times$ & 7.4$\times$\\\hline
$2^{13}$ & $2^{19}$ &  12.4 $ms$ &  0.93$\times$   & 15.46$\times$    &  7.36$\times$ & 11.1$\times$\\\hline
\end{tabular}%
}
\vspace{1ex}
\caption{Comparing the performance of multiplication and division with \cgbn{}. The first two columns indicate the characteristics of the dataset: the integer precision in number of bits and the batch size, i.e., number of instances that run in parallel. The third column shows the runtime of our multiplication. The fourth column shows the speedup of our multiplication {\em vs.} \cgbn{}. The fifth column shows the factor by which our division is slower than our multiplication. The sixth column shows the speedup of \cgbn{} division {\em vs.} our implementation of division; ``\textcolor{red}{$-$}'' indicates that \cgbn{} does not support that dataset. The last column shows the speedup of our \cuda{} division {\em vs.} the \textsc{GMP} division, run sequentially on an AMD EPYC 7352 CPU.} 
\label{tab:perf}
\end{table}
